\documentclass[12pt]{article}

\usepackage{amsmath}
\usepackage{amssymb}
\usepackage{graphicx}
\usepackage{booktabs}
\usepackage{geometry}
\usepackage{setspace}
\usepackage{natbib}
\usepackage{hyperref}
\usepackage{microtype}
\usepackage{float}
\usepackage{multirow}

\bibpunct{(}{)}{;}{a}{,}{,}

\title{\textbf{Pay-Per-Crawl Pricing for AI: The LM-Tree Agent}\thanks{We thank Dennis Bode for facilitating the research collaboration and providing access to HardwareLuxx data, and Rafael Biermann for his expert handling of the data pipeline. All errors are our own. Click \href{https://sites.google.com/view/soheil-ghili/ppc}{here} for the most current version. }}
\author{
  Richard Archer\thanks{Yale University. \texttt{richard.archer@yale.edu}.} \and
  Soheil Ghili\thanks{Yale University. \texttt{soheil.ghili@yale.edu}.} \and
  Nima Haghpanah\thanks{Yale University. \texttt{nima.haghpanah@yale.edu}.}
}
\date{\today}

\begin{document}

\maketitle

\begin{abstract}
As AI systems shift from directing users to content toward consuming it directly, publishers need a new revenue model: charging AI crawlers for content access. This model, called pay-per-crawl, must solve a problem of \emph{mechanism selection at scale}: content is too heterogeneous for a fixed pricing framework. Different sub-types warrant not only different price levels but different pricing rules based on different unstructured features, and there are too many to enumerate or design by hand. We propose the LM Tree, an adaptive pricing agent that grows a segmentation tree over the content library, using LLMs to discover what distinguishes high-value from low-value items and apply those attributes at scale, from binary purchase feedback alone. We evaluate the LM Tree on real content from a major German technology publisher, using 8,939 articles and 80,451 buyer queries with willingness-to-pay calibrated from actual AI crawler traffic. The LM Tree achieves a 65\% revenue gain over a single static price and a 47\% gain over two-category pricing, outperforming even the publisher's own 8-segment editorial taxonomy by 40\%---recovering content distinctions the publisher's own categories miss.
\end{abstract}

\newpage

\section{Introduction}

During the search era, the business model for online content was built on direct traffic. Users found content through search engines and accessed it directly; publishers monetized via advertising, subscriptions, or commerce. This model is eroding in the AI era. AI systems now crawl content for training and for retrieval-augmented generation, consuming publisher output without directing users to the source. The traffic-based revenue model breaks down.

For content creation to remain viable, publishers need a new revenue model: charging AI systems directly for content access. This is called \textbf{pay-per-crawl} (PPC). Services like Cloudflare and Tollbit now provide the infrastructure to charge bots. What remains unsolved is the pricing problem: how should a publisher determine what to charge?

Optimal pricing for PPC is hard for two reasons that compound each other. First, pricing-relevant features are unstructured: a content item's value to an AI crawler depends on features embedded in the text itself---topic specificity, data richness, timeliness---not on structured metadata. Pricing a piece of content requires reading and understanding it, not checking a category label. Second, the space of content types is massive and hierarchical: different content sub-types warrant different pricing rules based on different features, and the problem of deciding which features matter for which content must be solved at scale. Manual rule design is infeasible. Crucially, the relevant features differ not just within a publisher's catalog but across publisher types: for a financial news outlet, recency by the second determines value; for a legal database, the applicable jurisdiction is the key dimension; for a technology review site, the tier of the product under review drives willingness-to-pay. There is no universal pricing rule---each content type or subtype warrants its own discovered mechanism.

Together, these challenges constitute a problem of \emph{mechanism selection at scale}: a pricing methodology must discover--for any given publisher--which segments warrant distinct pricing, which textual features define those segments, and what prices they support---all from binary purchase feedback alone. The \textbf{LM Tree} is designed to solve it---an adaptive pricing agent that combines tree-based market segmentation with LLM-powered feature discovery. The agent starts with coarse content categories, explores prices, and observes only binary purchase outcomes. It deploys two specialized components: an \textbf{LLM Analyst} that reads content text and discovers what distinguishes high-value items from low-value ones---not from metadata, but from the prose itself---and an \textbf{LLM Annotator} that applies the discovered attributes to all items in the node. These attributes define split rules that partition content into finer sub-types, each priced independently. The process recurses, growing a pricing tree that matches the natural heterogeneity of the content. At inference time, no LLM is needed---split rules reduce to simple attribute lookups.

We evaluate the LM Tree on real content from HardwareLuxx (HWL), a major German technology publisher. The dataset contains 8,939 articles spanning two observable categories---long-form reviews (\textit{artikel}) and news items (\textit{news})---drawn from a corpus crawled extensively by AI systems including GPTBot, ClaudeBot, and PetalBot. Willingness-to-pay is calibrated from actual AI crawler traffic: $\text{WTP}(i) = 0.004 \times \text{observed crawler views}(i)$, reflecting the principle that crawlers' value for content scales with how actively they consume it in the current unpriced market. We generate 9 synthetic buyer queries per article (80,451 total) to simulate transaction-level demand. Articles are split into a training set (7,210 articles) on which the agent learns prices, and a test set (1,729 articles) on which learned prices are evaluated without further adjustment.

We train and compare a number of pricing policies. Each policy explores prices as our simulated queries ``arrive,'' receives a binary purchase feedback from each query, and optimizes the pricing of different content items over time. Importantly, the binary purchase feedback is the only information that each policy receives; the underlying calibrated WTP which determines the purchase decision is hidden from the agent.

Specifically, we have four pricing policies. Single-price optimization---one price for all 8,939 articles---yields \$160 in test-set revenue. Format category pricing---one price per content format, distinguishing long-form reviews (\textit{artikel}) from short news items (\textit{news})---raises this to \$179, a modest gain reflecting that this format distinction captures little of the within-format WTP heterogeneity. Moving to the publisher's own 8-segment editorial taxonomy (hardware, software, consumer electronics, and others within each format) reaches only \$189---an 18\% gain over single pricing. The LM Tree, initialized with only the two format categories and discovering finer content segments from article text alone, achieves \$264---a 65\% gain over single-price, a 47\% gain over format category pricing, and a 40\% gain over the richer 8-segment baseline.

The key finding concerns how the LM Tree structures its splits. The discovered rules cut across the publisher's formal editorial categories: an article covering high-end GPU specifications belongs to a different pricing regime than other hardware articles---a distinction the editorial taxonomy does not capture. The agent recovers this segmentation from binary purchase feedback alone, with no prior knowledge of the content hierarchy.

We expect pay-per-crawl to be the first but not the last market in which agentic pricing based on unstructured features becomes the dominant method. As AI systems intermediate more transactions---from API access to data licensing to professional services---the same core problem recurs: heterogeneous goods, unobservable willingness-to-pay, and pricing-relevant features buried in unstructured text. Pay-per-crawl provides an attractive setting in which to study the design of pricing agents supported by language models precisely because it combines all of these challenges in a market that is forming now.

The remainder of the paper proceeds as follows. Section~\ref{sec:lit} reviews related literature. Section~\ref{sec:setting} formalizes the setting: the publisher's content library, the buyer model, and the pricing problem. Section~\ref{sec:algorithm} presents the LM Tree---its algorithmic structure, the roles of the LLM Analyst and Annotator, and the properties of its splitting rules. Section~\ref{sec:data} describes the HardwareLuxx dataset and evaluation setup. Section~\ref{sec:results} reports results, including a decomposition of the revenue gains and an analysis of the splits the LM Tree makes. Section~\ref{sec:discussion} discusses broader applicability and open questions. Section~\ref{sec:conclusion} concludes.

\section{Related Literature}
\label{sec:lit}

\textbf{Bandit pricing and demand learning.} The price exploration step at each node is a multi-armed bandit problem. A substantial literature characterizes the cost of learning demand from binary purchase feedback: \citet{kleinberg2003} establish regret bounds for online posted-price auctions; \citet{besbes2009} derive near-optimal algorithms for dynamic pricing without prior knowledge of the demand function; \citet{denboer2015} surveys the subsequent development of these ideas across pricing contexts. In marketing, \citet{misra2019} demonstrate the practical value of bandit experiments for dynamic online pricing in field settings, showing that adaptive price exploration yields substantial revenue gains over passive approaches. The LM Tree applies this machinery locally---one bandit problem per node---but its primary contribution is not to the price-learning step. What it adds is a principled method for deciding \emph{which} bandit problems to run: the tree structure determines which items should be priced together, and the LLM Analyst determines when and how to split. Standard bandit pricing takes the set of items as given; the LM Tree discovers the right partition from data.

\textbf{Tree-based market segmentation.} Recursive partitioning methods have a long history in statistics \citep{breiman1984} and have been adapted to a range of causal and economic objectives. \citet{athey2016} introduce causal trees, which tailor split criteria to maximize treatment effect heterogeneity rather than prediction accuracy---establishing the principle that tree structure should be designed around the quantity of interest. \citet{wager2018} extend this to forests for valid nonparametric inference. \citet{aouad2023} bring the framework explicitly to pricing, developing market segmentation trees that split on features to maximize revenue differentiation across leaves. The LM Tree builds on this lineage and follows it closely in structure: the splitting criterion targets price differentiation, split validation uses economic rather than statistical criteria, and leaf prices are estimated from fresh exploration data. The departure is upstream: in all existing tree methods, the feature space is fixed and given---the algorithm's task is feature \emph{selection}, searching over columns of $X$. The LM Tree adds a prior step of feature \emph{construction}, in which the LLM Analyst generates a node-specific feature representation from content text. This is what allows the tree to operate in settings where no feature matrix exists.

\textbf{Text as data and LLM-based feature extraction.} A growing literature in economics uses text to measure quantities that are difficult to observe directly---\citet{gentzkow2019} survey applications ranging from political polarization to asset pricing. More recent work uses large language models for structured information extraction: given unstructured prose, an LLM can reliably extract attributes, labels, and relational facts that would otherwise require costly manual annotation. The LM Tree embeds this capability inside an adaptive economic loop. The LLM is not merely measuring a fixed target---it is discovering, at each node, which attributes are worth measuring, and the tree structure determines which discovery problems to pose. The interaction between economic feedback ($H$ and $L$ sets) and language model reasoning is the core mechanism; neither is useful without the other.

\textbf{Economics of AI and digital content markets.} The pay-per-crawl setting connects to a broader literature on the economics of data and AI. \citet{bergemann2015} study how data generates value in markets with asymmetric information; \citet{agarwal2018} provide a framework for thinking about AI as a prediction technology with implications for market structure. More directly, a nascent literature examines the economic relationship between AI developers and content producers---who owns training data, how scraping affects incentives to create, and what compensation mechanisms are feasible. The LM Tree takes the infrastructure of pay-per-crawl as given and asks how a publisher should price within it---a question this literature has not yet addressed formally.

\textbf{Price discrimination and product design.} At the economic level, the LM Tree is performing second-degree price discrimination: it partitions a heterogeneous product space into segments and prices each separately. The classical treatments---\citet{mussa1978} on monopoly and product quality, and \citet{maskin1984} on monopoly with incomplete information---assume the seller knows which product dimensions matter and designs quality schedules accordingly; \citet{ghili2025} combine theory, experimental design, and empirical estimation to study second-degree price discrimination in a unified framework. \citet{bergemann2015limits} characterize the full set of consumer surplus and profit outcomes achievable under any market segmentation, establishing the theoretical boundaries of price discrimination as an information design problem. \citet{haghpanah2022} study the limits of multiproduct price discrimination; \citet{haghpanah2023} identify conditions under which segmentation Pareto-improves market outcomes. On bundling---an extension of the second-degree framework to settings where the relevant quality dimensions interact---\citet{ghili2023} characterizes optimal bundling with nonadditive values; \citet{haghpanah2021} characterize when pure bundling is optimal; and \citet{yang2025} studies nested bundling, showing that hierarchical bundling structures can replicate complex nonlinear pricing schemes. \citet{shapiro1999} establish versioning as the canonical digital-goods strategy. \citet{dube2023} study personalized pricing---third-degree discrimination based on buyer identity---implemented at scale via machine learning, demonstrating large profit gains from buyer-level price targeting. The LM Tree as presented conditions prices on content characteristics rather than buyer identity, but extends naturally to a combination of second- and third-degree discrimination once buyer identity is observable; we discuss this in Section~\ref{sec:discussion}. All of this literature assumes the seller knows which product dimensions matter. The LM Tree relaxes this assumption: it discovers the relevant dimensions from data, making price discrimination applicable when the product space is too complex, heterogeneous, or novel to enumerate in advance.

\section{Setting}
\label{sec:setting}

A publisher operates a content library consisting of items $i \in \mathcal{I}$. Each item belongs to an observable coarse category $c(i) \in \mathcal{C}$ (e.g., hardware reviews, software news, consumer electronics coverage) and has associated text $t_i$---the prose body of the content. The text is the publisher's primary asset: it is what AI crawlers seek to access.

Buyers are AI crawlers operated by AI companies. Each buyer arrival $q$ targets a content category and is characterized by a buyer type $s_q$ (e.g., a large foundation model trainer vs.\ a retrieval-augmented generation service). Upon arrival, the buyer requests a content item $i$ from the publisher, who offers access at price $p$.

\subsection{Willingness-to-Pay}

Each buyer has a willingness-to-pay $v(i, s_q)$ that depends on the content item and the buyer type. What makes the pricing problem difficult is the structure of $v$: it is not a simple function of a few universal features, but a deeply heterogeneous object whose relevant dimensions change across content types.

Consider a publisher whose library spans several broad categories---say, GPU and processor benchmark reviews, software and games coverage, and consumer electronics articles. Within GPU benchmarks, the value of a piece of content to an AI training pipeline depends on rasterization throughput and thermal performance under sustained load. Within software and games coverage, these metrics are irrelevant; value depends on benchmark scores, frame rates, and driver compatibility specifics. Within consumer electronics, value depends on hands-on testing depth and product availability at time of publication.

These are not different values of the same attributes---they are entirely different attributes, measured in different units, relevant to different buyer types. A pricing rule that works for GPU benchmarks (e.g., ``charge more when silicon generation is current'') is meaningless for software news, where silicon generation is not a concept. Conversely, ``charge more for benchmark score coverage'' is meaningless for consumer electronics reviews.

The heterogeneity compounds as one looks deeper. Within GPU benchmarks, there are sub-types: flagship performance reviews of the latest silicon, budget-tier value comparisons, and legacy architecture retrospectives. Each commands different prices from different crawler types: a model trainer values cutting-edge silicon specs; a product recommendation engine values budget comparisons; a historical analysis pipeline values legacy coverage. The relevant pricing dimensions shift again at each level of granularity.

The result is that $v$ is not one function but a large family of functions, each defined over a different feature space, applying to a different slice of the content library. The number of distinct pricing rules one would need to write---each specifying which attributes matter, how they enter $v$, and what price they justify---grows combinatorially with the depth of the content taxonomy. For a publisher with thousands of content items across dozens of sub-types, manually specifying these rules is infeasible.

\subsection{Transactions}

When the publisher offers item $i$ to buyer $q$ at price $p$, the buyer purchases if and only if the price is below their willingness-to-pay:
\[
  y = \mathbf{1}[p \leq v(i, s_q)]
\]
The publisher observes the binary outcome $y$ but never the value $v(i, s_q)$, the buyer type $s_q$, or which features of the content drove the buyer's valuation. The publisher does observe its own content---the text $t_i$ and category $c(i)$---and the history of prices offered and outcomes received.

\subsection{The Publisher's Problem}

The publisher chooses a pricing policy $\pi: \mathcal{I} \to \mathbb{R}_+$ mapping content items to prices. The objective is to maximize expected revenue:
\[
  \max_\pi \; \mathbb{E}\!\left[\sum_q p_{\pi}(i_q) \cdot \mathbf{1}[p_{\pi}(i_q) \leq v(i_q, s_q)]\right]
\]
where $i_q$ is the item matched to query $q$ and $p_\pi(i_q)$ is the price assigned by policy $\pi$.

In principle, this is a standard dynamic pricing problem: explore prices, estimate demand, and converge to revenue-maximizing prices. But the structure of $v$ described above transforms it into something harder. Because the value-relevant features differ across content types---and because the publisher does not know \emph{which} features matter for \emph{which} content---the publisher cannot simply estimate a demand curve. It must first discover the right segmentation: which items should be priced together, and on what basis they should be differentiated.

This is a problem of \textbf{mechanism selection at scale}. The publisher is not choosing one pricing rule; it is choosing a \emph{collection} of pricing rules, one for each content segment, where the segments themselves must be discovered. For each segment, the publisher must identify the relevant features, learn how those features relate to willingness-to-pay, and set prices accordingly---all from binary purchase feedback alone. The combinatorial structure of $v$ means that the space of possible segmentations is vast, and the cost of exploring the wrong segmentation is real revenue forgone.

The pricing agent's task, then, is to navigate this space efficiently: discover which content segments warrant distinct pricing, identify the textual features that define those segments, and learn per-segment prices---using only the content text $t_i$ and the history of purchase outcomes as inputs.

\paragraph{Instantiation on HardwareLuxx.} We implement and evaluate this problem on data from HardwareLuxx (HWL), a major German technology publisher. The content library consists of 8,939 articles in two observable categories---long-form reviews (\textit{artikel}) and news items (\textit{news}). Willingness-to-pay is calibrated from real AI crawler traffic as $v(i) = 0.004 \times \text{observed views}(i)$, reflecting the assumption that crawlers' current consumption reveals their relative valuations. The publisher observes each article's text and its coarse category, but never the crawler view counts. Articles are split into a training set (7,210) on which the agent learns, and a test set (1,729) on which learned prices are evaluated without further adjustment.

\section{The LM Tree}
\label{sec:algorithm}

This section develops the LM Tree, a pricing agent that solves the mechanism selection at scale problem defined in Section~\ref{sec:setting}. The agent must simultaneously discover content segments, identify the textual features that define them, and learn per-segment prices---all from binary purchase feedback.

\subsection{Tree-Based Pricing Policies}

We begin with the standard framework for tree-based partitioning \citep{breiman1984}. A pricing tree $\mathcal{T}$ is a binary tree that induces a partition $\Pi = \{\mathcal{I}_1, \ldots, \mathcal{I}_K\}$ of the content library $\mathcal{I}$ into $K$ segments, with each segment $k$ assigned a price $p_k$. The pricing policy maps each item to the price of its segment:
\[
  \pi_\mathcal{T}(i) = p_{\ell(i)}
\]
where $\ell(i)$ is the leaf segment containing item $i$.

In a standard decision tree, each internal node $n$ holds a split rule $\sigma_n$ defined over a pre-specified feature vector $x_i \in \mathbb{R}^d$: the algorithm searches over all features $j \in \{1, \ldots, d\}$ and all thresholds $\tau$ to find the split $(j, \tau)$ that optimizes a criterion---Gini impurity for classification \citep{breiman1984}, treatment effect heterogeneity for causal inference \citep{athey2016}, or response model fit for market segmentation \citep{aouad2023}.

The publisher's problem differs from all of these in a fundamental way: Content is unstructured, hencethe feature vector $x_i$ does not exist. The publisher has content text $t_i$ and a coarse category label $c(i)$, but no structured feature matrix. The attributes that matter for pricing---benchmark depth, jurisdictional coverage, signal freshness---are embedded in the prose, vary by content type, and are not known in advance. There is no column to split on until someone reads the text and decides what to extract.

The LM Tree addresses this by adding a \textbf{feature construction} step before the standard split-and-estimate logic. At each node, a large language model reads the content text, discovers pricing-relevant attributes, and constructs a local feature representation. The tree then splits on these constructed features using standard logic. The key departure from the tree literature is that the feature space is not given---it is generated, and it is different at every node.

\subsection{Algorithm}

The LM Tree grows by alternating between three operations at each node: \textbf{price exploration}, \textbf{feature discovery}, and \textbf{split validation}. We describe each in turn. 

\textbf{Initialization.} The tree begins with $|\mathcal{C}|$ root nodes, one per coarse category. Each root contains all items in its category: $\mathcal{I}_n = \{i : c(i) = c\}$ for $c \in \mathcal{C}$. In the HardwareLuxx evaluation, $|\mathcal{C}| = 2$, corresponding to long-form reviews (\textit{artikel}) and news items (\textit{news}).

\textbf{Price exploration.} At node $n$, the agent explores $K$ price arms $\{p_1, \ldots, p_K\}$ log-spaced around a baseline. For each arm $k$, it runs $M$ trials: a query $q$ arrives, an item $i \in \mathcal{I}_n$ is offered at $p_k$, and the agent observes $y = \mathbf{1}[p_k \leq v(i, s_q)]$. The agent estimates per-arm conversion rates and revenue:
\[
  \hat{r}_k = \frac{1}{M}\sum_{m=1}^{M} y_{k,m}, \qquad \hat{R}_k = p_k \cdot \hat{r}_k
\]
where $\hat{r}_k$ is the empirical conversion rate (fraction of trials resulting in a purchase) and $\hat{R}_k$ is the estimated revenue per query. The node's current optimal price is $p^*_n = \arg\max_k \hat{R}_k$.

This is the standard multi-armed bandit approach to price discovery \citep{kleinberg2003}. Used alone, it yields a single price per category---the Category Pricing benchmark in our results. The LM Tree's contribution begins with what happens next.

\textbf{Feature discovery.} The goal is to identify which textual features of content drive high versus low willingness-to-pay. Two challenges make this hard. First, content is unstructured: pricing-relevant features are not in a column of a table but embedded in prose, and are not known in advance. Second, WTP is never directly observed---the agent has only binary purchase outcomes under different prices, and must infer latent value from this indirect signal.

The law of demand structures what can be inferred. High-value items sell at both high and low prices; low-value items sell only at low prices. As a result, the set of items that sell at high prices is a \emph{subset} of the set that sells at low prices---not a disjoint group. The agent's task is to find the hidden textual features that determine whether an item belongs to this high-value subset, using only the content text and the pattern of purchase outcomes under price variation.

To construct contrast sets that expose this structure, the agent partitions the $K$ price arms into a top half and a bottom half by rank:
\begin{itemize}
  \item $H_n$: items that purchased at a \emph{top-half} price arm---revealed willingness to pay at high prices
  \item $L_n$: items that purchased at a \emph{bottom-half} price arm\footnote{If $K$ is odd, the middle arm is assigned to the top half, erring toward a lower H threshold and thus more H observations. If $H_n$ remains too small after this partitioning---because high-price arms rarely convert---the highest-ranked bottom-half arm is iteratively reassigned to the top half until $H_n$ reaches a usable size.} ---revealed willingness to pay only at low prices
\end{itemize}

In a standard tree, the algorithm would search over columns of a feature matrix $X$ to find the split that best separates $H_n$ from $L_n$. The LM Tree replaces this search with a call to the \textbf{LLM Analyst}---an LLM whose role is to reason over content text and surface pricing-relevant structure.

The LLM Analyst is presented with a sample of items from $H_n$ and $L_n$---specifically, their content text $t_i$---and asked: \emph{what textual attributes distinguish items that sell at high prices from those that sell only at low prices?} It reads the prose and returns a set of candidate attributes $\{a_1, \ldots, a_J\}$.

This is the step that has no analogue in the existing tree literature. The LLM Analyst is performing feature construction: it examines unstructured text and proposes a structured representation that the tree can split on. The attributes it discovers are specific to the node---a different node, covering different content, will engage the LLM Analyst anew and may yield entirely different attributes.

\textbf{Split rules.} The discovered attributes define a split rule $\sigma_n: \mathcal{I}_n \to \{L, R\}$. The LM Tree supports two types, tried in order of preference.

\emph{Existence rules.} The LLM identifies an attribute $a$ that is \emph{mentioned} in items from $H_n$ but \emph{absent} from items in $L_n$ (or vice versa). The split rule is:
\[
  \sigma_n(i) = \begin{cases} R & \text{if attribute } a \text{ is mentioned in } t_i \\ L & \text{otherwise} \end{cases}
\]
For example, the LLM might discover that high-value technology reviews mention ``engagement lift'' while low-value ones describe ``data points per decision.'' The split is not on the \emph{value} of engagement lift but on whether the concept appears at all.

\emph{Threshold rules (fallback).} If both groups mention the same attribute but at different magnitudes, the LLM proposes a numeric threshold: $\sigma_n(i) = \mathbf{1}[a(i) > \tau]$.

Existence rules are preferred because they are robust to the heterogeneity described in Section~\ref{sec:setting}: different content sub-types discuss fundamentally different \emph{kinds} of metrics, making presence/absence a stronger signal than numeric comparison across incommensurable scales.

\textbf{Annotation.} Once the LLM Analyst has identified the relevant attributes, a second component---the \textbf{LLM Annotator}---applies them to \emph{all} items in $\mathcal{I}_n$, not just the $H_n$ and $L_n$ samples. For each item $i$, the LLM Annotator extracts the discovered attributes from $t_i$, producing a local feature vector $\hat{a}_n(i)$. The split rule $\sigma_n$ then operates on $\hat{a}_n(i)$ via a simple lookup. No LLM call is needed at pricing time.

Note the parallel to honest estimation in causal trees \citep{athey2016}: the exploration data used to discover the split ($H_n$ and $L_n$) is distinct from the data used to estimate prices in the child nodes (fresh exploration in children). This separation prevents overfitting the segmentation to noise in a particular set of purchase outcomes.

\textbf{Split validation.} After splitting node $n$ into children $n_L$ and $n_R$, the agent runs price exploration in each child independently. The split is retained if and only if the children's optimal prices differ: $p^*_{n_L} \neq p^*_{n_R}$. If both children converge to the same price, the split is discarded---it may separate items along a textually salient dimension, but not one that is economically relevant.

\textbf{Recursion.} At the conclusion of the procedure above, node $n$ is in one of two states. If no valid split was found---either the LLM Analyst did not surface a useful attribute or split validation failed because the two children converged to the same price---the node remains intact as a leaf, assigned price $p^*_n$, and the LM Tree is finished with it. If a valid split was found, node $n$ is partitioned into two child nodes $n'$ and $n''$, each containing its respective subset of content items and each carrying its own discovered price. For each child, the entire procedure---price exploration, feature discovery, annotation, and split validation---may be repeated. The tree grows recursively in this way, but will not recurse beyond a maximum depth $D$. Nodes at depth $D$ are always treated as leaves regardless of whether a further split might be beneficial.

\begin{figure}[H]
  \centering
  \includegraphics[width=0.85\textwidth]{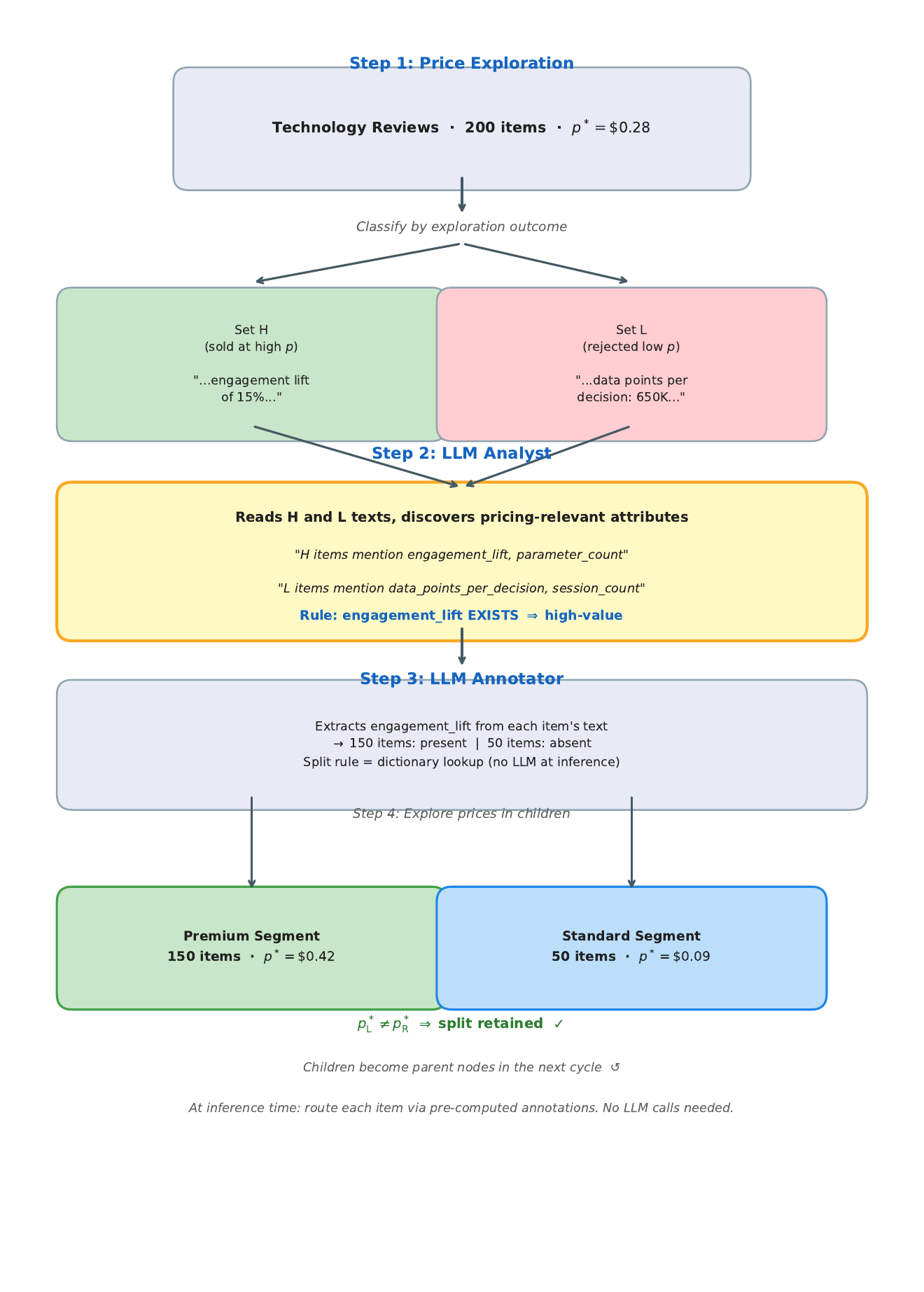}
  \caption{One cycle of the LM Tree at a single node. The LLM Analyst reads content from sets $H$ and $L$ to discover a pricing-relevant split rule; the LLM Annotator applies that rule to all items in the node. At inference time, routing requires only the pre-computed annotations---no LLM calls.}
  \label{fig:schematic}
\end{figure}

Figure~\ref{fig:schematic} illustrates one complete cycle at a single node.

\subsection{Algorithmic Properties}

Two design choices deserve attention beyond the core algorithmic structure.

\textbf{Log-scale exploration anchored to the parent's optimal price.} These two choices operate as a unit. At each node $n$, the price arms are spaced on a log scale---multiplicatively symmetric around a baseline, so that a $10\times$ upward exploration mirrors a $10\times$ downward exploration. This is the natural geometry for prices, where ratios matter more than absolute differences. The baseline is not fixed globally but inherited from the parent node's optimal price $p^*_{\text{parent}}$, which has already been validated one level up the tree.

Together, these choices allow the agent to approximate heterogeneous optimal prices even when they differ by orders of magnitude across content types. A publisher's catalog may contain commodity news items worth fractions of a cent and flagship benchmark reviews worth several dollars---a range spanning two or more orders of magnitude. The agent does not need to know this range in advance. At the root, it explores around a rough initial baseline. At each split, child nodes inherit a refined anchor from their parent and re-explore on a log scale centered there. Over successive rounds of recursion, each branch converges toward the true optimal price for its content sub-type, regardless of where that price sits in absolute terms.

\textbf{Price exploration plays a dual role.} The exploration trials at node $n$ serve two distinct purposes across two different rounds of the recursion. In the \emph{current} round, they identify the best price for node $n$: by spreading trials across a wide log-spaced range, the agent estimates per-arm revenue and selects $p^*_n$. In the \emph{next} round, the same trials generate the variation necessary to discover the right split rule and create child nodes $n'$ and $n''$: because prices are spread wide, some items reveal high WTP (they sell at top-half arms) and others reveal low WTP (they fail at bottom-half arms), producing the $H_n$ and $L_n$ contrast sets the LLM Analyst needs. The variation required for segmentation in the next round is a byproduct of the variation required for pricing in the current round, at no additional data collection cost.

With the agent described, we next turn to our data.

\section{Data: HardwareLuxx}
\label{sec:data}

We evaluate the LM Tree on real content from HardwareLuxx (HWL), a major German technology publisher whose editorial archive is crawled extensively by AI systems. The dataset combines two components: observed AI crawler traffic (real data from HWL's server logs) and calibrated willingness-to-pay (a constructed proxy, described below). We keep these distinct throughout.

\subsection{Traffic Data}

HardwareLuxx records AI crawler activity via Cloudflare, capturing requests from major AI systems including GPTBot (OpenAI training), OAI-SearchBot (OpenAI retrieval), ClaudeBot (Anthropic), and PetalBot (Huawei). Over a 30-day window, these crawlers generated hundreds of thousands of requests across HWL's editorial archive, with access rates varying substantially across articles and consistent with a power-law distribution.

HWL's editorial content is organized in a three-level taxonomy derived from its Joomla CMS. The two observable \textbf{format categories} constitute the coarse partition $\mathcal{C}$ available to the agent: long-form reviews and benchmark articles (\textit{artikel}) and shorter news items (\textit{news}). This format distinction is coarse---both categories span hardware, software, and consumer electronics coverage---and captures article style rather than content. The finer \textbf{editorial categories} (eight content domains such as hardware, software, and consumer electronics, nested within each format) are \emph{not} observable to the agent; they must be discovered from text.

We use 8,939 articles for which full text is available (content published from mid-2019 onward). Articles are split into a training set of 7,210 articles, on which the agent learns prices, and a test set of 1,729 articles, on which learned prices are applied without further adjustment. The split is stratified by format category and is across articles: no article appears in both sets.

\begin{table}[t]
\centering
\caption{AI crawler views by format category}
\label{tab:hwl_categories}
\small
\begin{tabular}{lrrrrr}
\toprule
Category & Articles & Median & Mean & Std & CV \\
\midrule
\textit{Artikel} (reviews) & 1,624 & 15.0 & 22.8 & 26.2 & 1.15 \\
\textit{News}              & 7,315 &  4.0 &  5.8 &  4.8 & 0.81 \\
\midrule
Total                      & 8,939 &  5.0 &  8.9 & 13.6 & 1.53 \\
\bottomrule
\end{tabular}
\end{table}

Reviews attract substantially more AI crawler attention per article than news items (median 15 vs.\ 4 views), a 3.8$\times$ gap that reflects that long-form benchmark content is more intensively consumed by AI systems than short product announcements. Traffic is also highly concentrated: the top 10\% of articles account for 42\% of total crawler requests, and the top 25\% account for 64\%. This concentration motivates the tree structure---a flat pricing policy leaves substantial revenue on the table by treating the high-traffic tail the same as the long tail of lightly-crawled articles.

Table~\ref{tab:hwl_l1} reports crawler-view statistics by editorial category---the eight content domains beneath the two observable format categories.

\begin{table}[t]
\centering
\caption{AI crawler views by editorial category}
\label{tab:hwl_l1}
\small
\begin{tabular}{llrrrrr}
\toprule
Format & Editorial category & Articles & Median & Mean & Std & CV \\
\midrule
\textit{Artikel} & Hardware             & 1,371 & 17.0 & 25.0 & 27.5 & 1.10 \\
\textit{Artikel} & Software             &   124 &  6.0 &  6.9 &  4.9 & 0.71 \\
\textit{Artikel} & Consumer electronics &   111 &  9.0 & 15.0 & 14.9 & 0.99 \\
\textit{Artikel} & Miscellaneous        &    18 &  7.0 &  8.7 &  6.7 & 0.77 \\
\midrule
\textit{News}    & Hardware             & 3,744 &  4.0 &  6.0 &  4.8 & 0.80 \\
\textit{News}    & Software             & 1,663 &  4.0 &  5.4 &  4.7 & 0.87 \\
\textit{News}    & General              & 1,209 &  4.0 &  5.4 &  4.5 & 0.83 \\
\textit{News}    & Consumer electronics &   699 &  5.0 &  6.5 &  4.7 & 0.73 \\
\bottomrule
\end{tabular}
\end{table}

\paragraph{Interpretation.} Three patterns in Table~\ref{tab:hwl_l1} motivate the LM Tree design. First, the format split does meaningful work: \textit{artikel} hardware (median 17) towers over every news sub-category (median 4--5), a gap the agent can exploit with a single coarse partition. Second, within \textit{artikel}, hardware is a clear outlier---the remaining three sub-categories (software, consumer electronics, miscellaneous) have medians of 6--9, closer to one another than to hardware, so the relevant within-\textit{artikel} split is ``hardware vs.\ rest,'' not a smooth gradient across all four domains. Third, within \textit{news}, the four sub-categories are nearly indistinguishable by median (all 4--5) and by CV (0.73--0.87): the editorial taxonomy adds no pricing information for news items. Whatever variation exists within news is article-level, not sub-category-level---it must be recovered from text, not from the category label. This is precisely the setting where the LM Tree's text-based splits are most valuable.

\paragraph{Crawler-level heterogeneity.} The five crawlers in Table~\ref{tab:hwl_bots} differ substantially in the content they seek. GPTBot directs 85\% of its requests to short news items, consistent with broad-coverage training data collection. OAI-SearchBot---from the same parent company---does the opposite: 73\% of its requests target long-form \textit{artikel} reviews, consistent with retrieval for search grounding where authoritative depth matters more than breadth. PetalBot, ClaudeBot, and PerplexityBot occupy intermediate positions, with a roughly even split between the two formats.

\begin{table}[t]
\centering
\caption{AI crawler requests by bot and content format (30-day window)}
\label{tab:hwl_bots}
\small
\begin{tabular}{lrrr}
\toprule
Bot & Total requests & \% \textit{Artikel} & \% \textit{News} \\
\midrule
PetalBot       & 468,811 & 44.5 & 55.5 \\
GPTBot         & 168,370 & 14.7 & 85.3 \\
OAI-SearchBot  &  42,718 & 73.0 & 27.0 \\
ClaudeBot      &  13,044 & 41.7 & 58.3 \\
PerplexityBot  &   9,077 & 45.7 & 54.3 \\
\bottomrule
\end{tabular}
\end{table}

This heterogeneity implies that the same article carries different value to different crawlers---the \textit{artikel} hardware review that OAI-SearchBot heavily targets is of little interest to a GPTBot sweep of news items. Optimal pricing would therefore combine second-degree price discrimination (segmenting by content characteristics, as the LM Tree does) with third-degree price discrimination (setting different prices for different crawler identities). Crawler identity is observable in practice: AI systems authenticate when accessing gated content, making buyer type a feasible conditioning variable. The LM Tree framework is readily extensible in this direction---the Level-0 partition $\mathcal{C}$ can be defined over (content category $\times$ crawler identity) rather than content category alone. In this paper we focus on the content dimension only, leaving the joint segmentation for future work.

\subsection{WTP Calibration}

No pay-per-crawl pricing is currently live at HardwareLuxx---as is typical of the market at this stage. The infrastructure to charge AI crawlers per access (HTTP 402, Tollbit, Cloudflare's bot management layer) is only now being adopted by early-moving publishers \citep{stackoverflow2026}. Transaction-level willingness-to-pay data does not yet exist for HWL or for most publishers. We therefore calibrate WTP from the traffic data:
\[
  v(i) = 0.004 \times \text{observed AI crawler views}(i)
\]
The coefficient $0.004$ is chosen so that the median article WTP is \$0.02---a plausible per-access price for the nascent pay-per-crawl market, consistent with the range discussed in early industry implementations. Table~\ref{tab:hwl_wtp} reports the resulting distribution.

\begin{table}[t]
\centering
\caption{Calibrated WTP distribution ($v(i) = 0.004 \times \text{AI views}$)}
\label{tab:hwl_wtp}
\begin{tabular}{lcccc}
\toprule
 & Min & Median & Mean & Max \\
\midrule
All articles & \$0.004 & \$0.020 & \$0.036 & \$1.112 \\
\textit{Artikel} (reviews) & \$0.004 & \$0.060 & \$0.091 & \$1.112 \\
\textit{News} & \$0.004 & \$0.016 & \$0.023 & \$0.160 \\
\bottomrule
\end{tabular}
\end{table}

Two arguments justify this calibration approach. The first is directional: in an unpriced market, AI systems allocate their crawl budget toward content they find most useful, so access frequency is a plausible proxy for value. Crawlers visit more what they would pay more for. The second argument is structural and more fundamental. Even if traffic is an imperfect directional proxy for WTP, the traffic data has exactly the statistical structure that a real WTP distribution would need to have in order to test the LM Tree's core claim. As Tables~\ref{tab:hwl_categories}--\ref{tab:hwl_l1} show, crawler views are partially correlated with observable content categories---reviews attract more traffic than news, and hardware reviews attract more than software reviews---but category alone leaves substantial within-group variation unexplained. That residual variation is not random noise: the top articles within each category are identifiable from their content (flagship GPU benchmark reviews, canonical CPU comparisons), suggesting that the within-category signal is recoverable from text. A WTP distribution with this structure---coarse-category signal plus large text-recoverable within-category variation---is precisely the setting in which the LM Tree is designed to add value over a category-level pricer. By calibrating WTP from traffic, we construct a test environment that is structurally valid for our evaluation regardless of whether traffic tracks WTP levels precisely.

\section{Agent Training}
\label{sec:training}

This section describes how the pricing agents are trained on the HardwareLuxx data. We first explain how buyer queries are generated to simulate transaction-level demand, then describe the market simulation environment in which agents learn prices from binary purchase feedback.

\subsection{Query Generation}

Each article receives 9 synthetic buyer queries, yielding 80,451 queries in total (64,890 training, 15,561 test). Query-level WTP is drawn independently as:
\[
  v(i, q) \sim \mathcal{N}\!\left(v(i),\; \sigma^2\right), \quad \sigma = 0.001, \quad \text{truncated at } 0
\]
The small $\sigma$ captures idiosyncratic variation across requests while keeping WTP close to the article-level center. Arrival order is randomized across all articles so the agent encounters content in a shuffled sequence rather than in batches by category.

\subsection{Market Simulation and Online Training of Agents}

We train three pricing agents of increasing sophistication on 8,939 real HardwareLuxx articles (7,210 train, 1,729 test). All agents use the same log-scaled grid exploration with binary purchase feedback. Each agent observes each article's text $t_i$ and its coarse category $c(i) \in \{\textit{artikel},\, \textit{news}\}$, but never the crawler view counts or the query-level WTP. It receives only the binary purchase outcome after each pricing decision.

The three agents are:
\begin{itemize}
  \item \textbf{Single Price:} A single price for all content, discovered by log-spaced grid exploration over the full query stream.
  \item \textbf{Category Pricing:} One price per category, discovered independently by exploration within each category. We consider two variants: a coarse partition into the two format categories (artikel, news), and a finer partition into the eight editorial categories (artikel/hardware, artikel/software, news/hardware, etc.).
  \item \textbf{LM Tree:} The agent described in Section~\ref{sec:algorithm}. It begins with the two format categories and uses the LLM Analyst to discover finer segmentation from article text.
\end{itemize}

\section{Empirical Results}
\label{sec:results}

We evaluate the three agents on the held-out test set and examine the pricing rules they produce. Section~\ref{sec:perf} compares revenue across agents. Section~\ref{sec:splits} looks inside the learned prices and splits to understand what drives the gains.

\subsection{Performance Analysis}
\label{sec:perf}

Table~\ref{tab:revenue} reports revenue for each strategy on both the training and test sets. Percentage comparisons are based on test-set revenue.

\begin{table}[t]
\centering
\caption{Revenue by pricing strategy (HardwareLuxx; 7,210 train articles, 1,729 test articles). Percentage gains are computed on the test set.}
\label{tab:revenue}
\begin{tabular}{lcccc}
\toprule
Strategy & Train & Test & vs.\ Single Price & vs.\ Format (2) \\
\midrule
Single Price                  & \$690  & \$160 & ---      & ---     \\
Category Pricing -- Format (2) & \$726  & \$179 & $+$12\%  & ---     \\
Category Pricing (8 segments) & \$772  & \$189 & $+$18\%  & $+$6\%  \\
LM Tree                       & \$1,075 & \$264 & $+$65\%  & $+$47\% \\
\bottomrule
\end{tabular}
\end{table}

The format categories (artikel vs.\ news) are a \emph{format} distinction rather than a content distinction: both categories cover the same product domains, differing mainly in article depth and style. Accordingly, format category pricing gains only 12\% over a single price. Moving to the finer eight-segment editorial partition---which separates hardware from software from consumer electronics within each format---adds a further 6\%, reaching 18\% above the single-price baseline. The editorial taxonomy helps, but both category pricing variants leave the majority of WTP heterogeneity unexploited.

The LM Tree achieves \$264 on the test set, a 65\% gain over single pricing and a 47\% gain over format category pricing. Notably, the LM Tree is initialized with only the two format categories; it has no access to the eight-segment editorial taxonomy. Yet it outperforms editorial category pricing by 40\%. The splits the agent discovers---articles covering high-end GPU and CPU components, precise performance specifications, advanced cooling solutions---cut across the formal editorial boundaries in ways that align more closely with AI crawler WTP than the publisher's own content taxonomy.

The train--test pattern reflects the cost of online learning rather than overfitting. The LM Tree accumulates \$1,075 on the training stream, which includes the exploration cost of the log-scaled price grid; on the held-out test set, at the prices learned during training, it earns \$264. The gap is proportional across strategies: format category pricing shows a similar train/test ratio, confirming that the drop is a feature of the exploration protocol, not a property of the LM Tree specifically.

\subsection{Prices and Splits}
\label{sec:splits}

\subsubsection{Observable-Based Policies}

Table~\ref{tab:prices_obs} shows the prices chosen by each observable-based policy after training.

\begin{table}[t]
\centering
\caption{Learned prices for observable-based policies. Editorial category names follow the HardwareLuxx editorial taxonomy.}
\label{tab:prices_obs}
\begin{tabular}{llc}
\toprule
Policy & Segment & Price \\
\midrule
Single Price & All content & \$0.055 \\
\midrule
\multirow{2}{*}{Category Pricing -- Format (2)}
  & news    & \$0.028 \\
  & artikel & \$0.263 \\
\midrule
\multirow{8}{*}{Category Pricing (8 segments)}
  & artikel/hardware             & \$0.263 \\
  & artikel/consumer-electronics & \$0.045 \\
  & artikel/sonstiges            & \$0.034 \\
  & artikel/software             & \$0.030 \\
  & news/hardware                & \$0.028 \\
  & news/software                & \$0.028 \\
  & news/consumer-electronics    & \$0.028 \\
  & news/allgemein               & \$0.028 \\
\bottomrule
\end{tabular}
\end{table}

Moving from a single price to the two format categories reveals a sharp format premium: artikel is priced at \$0.263, nearly ten times the news price of \$0.028. This reflects the aggregate WTP difference between in-depth hardware reviews and news announcements. Moving to the eight-segment editorial partition breaks the artikel category apart: hardware reviews retain the high price (\$0.263), while consumer-electronics articles (\$0.045), miscellaneous articles (\$0.034), and software articles (\$0.030) are priced down closer to their true WTP centers. All four news sub-types converge to the same price (\$0.028), indicating that WTP is homogeneous within the news format regardless of topic. The finer partition thus enables more precise pricing for the heterogeneous artikel segment, which is where most of the revenue opportunity lies.

\subsubsection{The LM Tree}

Table~\ref{tab:prices_lmt} shows the four prices learned by the LM Tree.

\begin{table}[t]
\centering
\caption{Learned prices for the LM Tree. Each leaf corresponds to one split rule applied within a format category.}
\label{tab:prices_lmt}
\begin{tabular}{llp{6.5cm}c}
\toprule
Format category & Leaf & Split rule & Price \\
\midrule
\multirow{2}{*}{news}
  & high-value & Market value $\geq$ \$1,000 (threshold) & \$0.023 \\
  & low-value  & otherwise                                & \$0.019 \\
\midrule
\multirow{2}{*}{artikel}
  & high-value & High-end GPU mentioned (existence)       & \$0.148 \\
  & low-value  & otherwise                                & \$0.081 \\
\bottomrule
\end{tabular}
\end{table}

The full rule descriptions produced by the LLM Analyst are:
\begin{itemize}
  \item \textbf{news:} ``Select items that have a market value of at least 1000 USD, indicating they are likely to be high value content.''
  \item \textbf{artikel:} ``Identify high-performing content by checking the presence of high-end GPU specifications like NVIDIA GeForce RTX 30 series or AMD equivalent in the content. This attribute is likely to be present in high-value content but absent in low-value content.''
\end{itemize}

Two features of the learned tree stand out. First, the artikel split produces the largest price separation: GPU and CPU review articles are priced at \$0.148, while other hardware reviews are priced at \$0.081---both above the highest news price (\$0.023), consistent with the format premium. Second, the news split is a threshold rule (``market value $\geq$ \$1,000'') rather than an existence rule, separating news coverage of high-end equipment from commodity product announcements with a value benchmark.

The LM Tree's splits do not coincide with editorial category boundaries. Table~\ref{tab:split_shares} shows the share of queries assigned to the high-value leaf within each editorial category.

\begin{table}[t]
\centering
\caption{Share of queries assigned to the high-value leaf, by editorial category. No editorial category maps cleanly onto either leaf, confirming that the LM Tree's splits cut across the formal content taxonomy.}
\label{tab:split_shares}
\begin{tabular}{llc}
\toprule
Format & Editorial category & Share assigned high \\
\midrule
\multirow{4}{*}{news}
  & news/allgemein               & 16.7\% \\
  & news/hardware                & 11.1\% \\
  & news/consumer-electronics    &  9.2\% \\
  & news/software                &  1.4\% \\
\midrule
\multirow{4}{*}{artikel}
  & artikel/sonstiges            & 40.0\% \\
  & artikel/hardware             & 17.4\% \\
  & artikel/software             &  6.9\% \\
  & artikel/consumer-electronics &  5.3\% \\
\bottomrule
\end{tabular}
\end{table}

For the artikel split (high-end GPU rule), artikel/hardware has the highest share assigned high (17.4\%), as expected---but the rule also reaches into artikel/sonstiges (40\%, small sample) and artikel/software (6.9\%). For the news split (market-value threshold), news/allgemein tops the table at 16.7\%, ahead of news/hardware (11.1\%)---general news coverage of high-value products triggers the threshold rule more than dedicated hardware news. No single editorial category is cleanly mapped to either leaf.

This cross-cutting structure is precisely what allows the LM Tree to outperform editorial category pricing despite being initialized with only the two format categories. The editorial taxonomy groups articles by topic; the LM Tree groups them by the textual signals that predict AI crawler WTP. These two partitions are correlated---GPU reviews do dominate the high-value leaf---but they are not the same. The agent recovers a segmentation that the publisher's own content hierarchy does not provide.

\section{Discussion}
\label{sec:discussion}

\subsection{Alternative Business Models for AI Content Access}

The emergence of AI systems as content consumers raises a prior question: what business model should govern this market? Three candidates are in play. Pay-per-crawl is one; the others each have a natural precedent in adjacent markets, but neither transfers cleanly to this setting.

\textbf{Bulk licensing.} The most visible current model is negotiated bulk licensing: a lump-sum payment grants an AI company broad access to a publisher's archive. Google's reported \$60 million deal with Reddit \citep{googlereddit2024} is the most prominent example. Such deals work well when both parties are large and the content is broadly relevant---they avoid per-transaction negotiation costs and provide predictable revenue. But they do not scale to the long tail of publishers, and they price content in aggregate: high-value technical documentation and commodity news are bundled at the same rate. A platform brokering deals on behalf of thousands of small publishers would face the mechanism selection problem in its most acute form, with no market signal and no discovery mechanism.

\textbf{Auctions.} Online advertising established keyword auctions---most famously, sponsored search---as the dominant mechanism for pricing scarce digital attention. The case for auctions rests on genuine slot scarcity: each query generates one sponsored position, and bidding resolves the allocation efficiently. No analogous scarcity exists in the content-to-AI market. AI crawlers can access any piece of content independently and without displacing any other buyer; there is no slot to auction. The efficiency rationale that makes search auctions work does not carry over.

\textbf{Pay-per-crawl.} Per-access pricing avoids both limitations. It scales without upfront negotiation---access is priced automatically per request---and it can reflect heterogeneous content value at the item level, provided the pricing mechanism is sophisticated enough to discover those differences. The latter is the problem the LM Tree is designed to solve. \citet{rothschild2025} conjecture more broadly that micro-transaction-based pricing will become a natural architecture of the agentic economy, as AI agents transact at a volume and granularity that bulk deals cannot accommodate.

Early market evidence suggests pay-per-crawl is gaining traction. Cloudflare's AI Audit product and Tollbit now provide infrastructure for per-access charging, and publishers including Stack Overflow have begun gating AI crawler access under such arrangements \citep{stackoverflow2026}. The market is nascent, but the infrastructure is in place. Whether pay-per-crawl becomes a dominant model---alone or alongside bulk licensing for the largest publishers---will depend on whether pricing mechanisms can adapt to the heterogeneity of content. That is the gap this paper addresses.

\subsection{Beyond Pay-Per-Crawl}

The pay-per-crawl setting is the motivating application, but the LM Tree is designed for a problem class, not a single market. The relevant conditions are three: goods are heterogeneous, WTP is unobservable, and the features that drive value are embedded in the text of the goods themselves rather than in a structured metadata table.

\textbf{API access pricing.} AI companies and infrastructure providers increasingly sell access to capabilities---model inference, retrieval pipelines, specialized tools---at per-call or per-token prices. The value of a particular API endpoint depends on what the endpoint does, which is described in documentation prose, not in a structured schema. An LM Tree could partition the endpoint catalog by the \emph{kinds of tasks} buyers use each endpoint for, discovering that endpoints described in terms of ``reasoning over long contexts'' command different prices than those described in terms of ``fast single-turn classification''---without requiring the seller to specify these categories in advance.

\textbf{Data licensing.} Data vendors license datasets to buyers whose intended use determines their WTP. The LM Tree's architecture applies directly: explore prices, identify high- and low-WTP buyers, ask the LLM Analyst what the descriptions of high-revenue datasets have in common, annotate the catalog, and split.

\textbf{Professional and expert services.} Pricing for consulting engagements, legal research, or specialized reports is often set by negotiation precisely because the value-relevant features---scope, depth, the specific expertise mobilized---are embedded in the engagement description, not in a billing code. A tree that discovers, from historical outcomes, which textual features of engagements predict high client WTP provides a principled pricing floor for future proposals.

In each of these markets, the same structural insight applies: the LM Tree does not require the seller to know in advance which features of their offerings matter to which buyers. The existence-rules finding extends naturally to these settings: products that mention ``real-time'' are categorically different from products that do not---the presence of the concept signals a different use case, a different infrastructure cost, and a different buyer profile.

\section{Conclusion}
\label{sec:conclusion}

As AI systems shift from delivering search results to consuming content directly, publishers need a new revenue model---and a new pricing technology. Pay-per-crawl creates the market; the LM Tree provides the pricing agent. It grows a segmentation tree over the content library, using an LLM Analyst to discover pricing-relevant attributes from content text at each node and an LLM Annotator to apply them at scale. The result is a system that discovers both the right segmentation and the right prices from binary purchase feedback, with no prior knowledge of which content features matter or how they interact with buyer willingness-to-pay.

Evaluated on 8,939 real HardwareLuxx articles, the LM Tree achieves a 65\% revenue gain over a single static price and a 47\% gain over two-category pricing---outperforming even the publisher's own 8-segment editorial taxonomy by 40\%---using only article text and binary purchase feedback. The agent's discovered splits cut across the publisher's formal editorial categories, recovering a segmentation based on textual signals that aligns more closely with AI crawler willingness-to-pay than the editorial taxonomy does. Content covering high-end GPU and CPU components belongs to a different pricing regime than commodity news about the same products---a distinction the publisher's own category labels do not capture.

The deeper contribution is to the tree literature: the LM Tree replaces feature selection with feature construction, enabling tree-based pricing in markets where the relevant dimensions are unknown, content-specific, and too numerous to enumerate---markets that share the structure of mechanism selection at scale. The value is not only in finding the right prices but in discovering the right pricing rule. A financial news outlet needs recency-based pricing; a legal database needs jurisdiction-based pricing; a product review site needs tier-of-product pricing. The LM Tree discovers which dimensions of a publisher's content drive AI crawler willingness-to-pay---without requiring the publisher to know this in advance. This is not unique to pay-per-crawl. Any market in which goods are described in text and willingness-to-pay is unobservable---API access, data licensing, professional services---shares the same structure. As AI intermediates more transactions, agents that learn from language as well as from prices will become increasingly important.

\newpage
\bibliographystyle{plainnat}
\bibliography{references}

@book{breiman1984,
  author    = {Breiman, Leo and Friedman, Jerome H. and Olshen, Richard A. and Stone, Charles J.},
  title     = {Classification and Regression Trees},
  publisher = {Wadsworth},
  year      = {1984}
}

@article{athey2016,
  author  = {Athey, Susan and Imbens, Guido},
  title   = {Recursive Partitioning for Heterogeneous Causal Effects},
  journal = {Proceedings of the National Academy of Sciences},
  volume  = {113},
  number  = {27},
  pages   = {7353--7360},
  year    = {2016}
}

@article{wager2018,
  author  = {Wager, Stefan and Athey, Susan},
  title   = {Estimation and Inference of Heterogeneous Treatment Effects using Random Forests},
  journal = {Journal of the American Statistical Association},
  volume  = {113},
  number  = {523},
  pages   = {1228--1242},
  year    = {2018}
}

@article{aouad2023,
  author  = {Aouad, Ali and Bandi, Chaithanya and Randhawa, Ramandeep},
  title   = {Market Segmentation Trees},
  journal = {Manufacturing \& Service Operations Management},
  year    = {2023}
}

@inproceedings{kleinberg2003,
  author    = {Kleinberg, Robert and Leighton, Tom},
  title     = {The Value of Knowing a Demand Curve: Bounds on Regret for Online Posted-Price Auctions},
  booktitle = {Proceedings of the 44th Annual IEEE Symposium on Foundations of Computer Science},
  pages     = {594--605},
  year      = {2003}
}

@article{besbes2009,
  author  = {Besbes, Omar and Zeevi, Assaf},
  title   = {Dynamic Pricing Without Knowing the Demand Function: Risk Bounds and Near-Optimal Algorithms},
  journal = {Operations Research},
  volume  = {57},
  number  = {6},
  pages   = {1407--1420},
  year    = {2009}
}

@article{denboer2015,
  author  = {den Boer, Arnoud V.},
  title   = {Dynamic Pricing and Learning: Historical Origins, Current Research, and New Directions},
  journal = {Surveys in Operations Research and Management Science},
  volume  = {20},
  number  = {1},
  pages   = {1--18},
  year    = {2015}
}

@article{gentzkow2019,
  author  = {Gentzkow, Matthew and Kelly, Bryan and Taddy, Matt},
  title   = {Text as Data},
  journal = {Journal of Economic Literature},
  volume  = {57},
  number  = {3},
  pages   = {535--574},
  year    = {2019}
}

@article{bergemann2015,
  author  = {Bergemann, Dirk and Bonatti, Alessandro},
  title   = {Selling Cookies},
  journal = {American Economic Journal: Microeconomics},
  volume  = {7},
  number  = {3},
  pages   = {259--294},
  year    = {2015}
}

@book{agarwal2018,
  author    = {Agarwal, Ajay and Gans, Joshua and Goldfarb, Avi},
  title     = {Prediction Machines: The Simple Economics of Artificial Intelligence},
  publisher = {Harvard Business Review Press},
  year      = {2018}
}

@article{mussa1978,
  author  = {Mussa, Michael and Rosen, Sherwin},
  title   = {Monopoly and Product Quality},
  journal = {Journal of Economic Theory},
  volume  = {18},
  number  = {2},
  pages   = {301--317},
  year    = {1978}
}

@book{shapiro1999,
  author    = {Shapiro, Carl and Varian, Hal R.},
  title     = {Information Rules: A Strategic Guide to the Network Economy},
  publisher = {Harvard Business School Press},
  year      = {1999}
}

@article{maskin1984,
  author  = {Maskin, Eric and Riley, John},
  title   = {Monopoly with Incomplete Information},
  journal = {RAND Journal of Economics},
  volume  = {15},
  number  = {2},
  pages   = {171--196},
  year    = {1984}
}

@article{bergemann2015limits,
  author  = {Bergemann, Dirk and Brooks, Benjamin and Morris, Stephen},
  title   = {The Limits of Price Discrimination},
  journal = {American Economic Review},
  volume  = {105},
  number  = {3},
  pages   = {921--957},
  year    = {2015}
}

@article{haghpanah2022,
  author  = {Haghpanah, Nima and Siegel, Ron},
  title   = {The Limits of Multiproduct Price Discrimination},
  journal = {American Economic Review: Insights},
  volume  = {4},
  number  = {4},
  pages   = {443--458},
  year    = {2022}
}

@article{haghpanah2023,
  author  = {Haghpanah, Nima and Siegel, Ron},
  title   = {Pareto-Improving Segmentation of Multiproduct Markets},
  journal = {Journal of Political Economy},
  volume  = {131},
  number  = {6},
  pages   = {1546--1575},
  year    = {2023}
}

@article{ghili2023,
  author  = {Ghili, Soheil},
  title   = {A Characterization for Optimal Bundling of Products with Nonadditive Values},
  journal = {American Economic Review: Insights},
  volume  = {5},
  number  = {3},
  pages   = {311--326},
  year    = {2023}
}

@article{haghpanah2021,
  author  = {Haghpanah, Nima and Hartline, Jason D.},
  title   = {When Is Pure Bundling Optimal?},
  journal = {Review of Economic Studies},
  volume  = {88},
  number  = {3},
  pages   = {1127--1156},
  year    = {2021}
}

@article{yang2025,
  author  = {Yang, Frank},
  title   = {Nested Bundling},
  journal = {American Economic Review},
  volume  = {115},
  number  = {9},
  pages   = {2970--3013},
  year    = {2025}
}

@unpublished{ghili2025,
  author = {Ghili, Soheil and Sudhir, K. and Jain, Nitish and Garg, Ankur},
  title  = {Second-Degree Price Discrimination: Theoretical Analysis, Experiment Design, and Empirical Estimation},
  note   = {Working paper},
  year   = {2025}
}

@article{dube2023,
  author  = {Dub\'{e}, Jean-Pierre and Misra, Sanjog},
  title   = {Personalized Pricing and Consumer Welfare},
  journal = {Journal of Political Economy},
  volume  = {131},
  number  = {1},
  pages   = {131--189},
  year    = {2023}
}

@misc{rothschild2025,
  author       = {Rothschild, David M. and Mobius, Markus and Hofman, Jake M. and Dillon, Eleanor and Goldstein, Daniel G. and Immorlica, Nicole and Jaffe, Sonia and Lucier, Brendan and Slivkins, Aleksandrs and Vogel, Matthew},
  title        = {The Agentic Economy},
  year         = {2025},
  howpublished = {arXiv:2505.15799},
  note         = {Microsoft Research}
}

@misc{googlereddit2024,
  author       = {{The New York Times}},
  title        = {Reddit in \$60 Million Deal for Training {A.I.} Systems},
  year         = {2024},
  howpublished = {\url{https://www.nytimes.com/2024/02/22/technology/reddit-google-artificial-intelligence.html}},
  note         = {February 22, 2024}
}

@misc{stackoverflow2026,
  author       = {{Stack Overflow}},
  title        = {Stack Overflow and {Cloudflare} Partner to Help Developers Protect and Monetize Content in the {AI} Era},
  year         = {2024},
  howpublished = {\url{https://stackoverflow.blog/2024/09/19/stack-overflow-and-cloudflare-partner-to-help-developers-protect-and-monetize-content-in-the-ai-era/}},
  note         = {September 2024}
}

@article{misra2019,
  author  = {Misra, Kanishka and Schwartz, Eric M. and Abernethy, Jacob},
  title   = {Dynamic Online Pricing with Incomplete Information Using Multiarmed Bandit Experiments},
  journal = {Marketing Science},
  volume  = {38},
  number  = {2},
  pages   = {226--252},
  year    = {2019}
}
\newpage
\section*{Appendices}
\appendix

\section{Data Preparation}
\label{app:data}

This appendix describes how the HardwareLuxx dataset was constructed from raw publisher data. The pipeline has four stages: (1) data extraction from the HWL content management system; (2) WTP calibration from AI-crawler traffic records; (3) train/test splitting; and (4) query generation. The pipeline is designed around the same invariant as the theoretical setting: the agent observes only the coarse category $c(i)$ and content text $t_i$ of each article. Crawler view counts, query-level WTP, and the full editorial taxonomy are hidden. The only feedback available to the agent is the binary purchase outcome $y = \mathbf{1}[p \leq v(i, q)]$.

\subsection{Data Extraction}
\label{app:extraction}

HardwareLuxx publishes articles through a Joomla content management system. The data was extracted via a direct database export coordinated with the HWL technical team. Two tables were used: the article metadata table (containing article ID, title, publication date, and Joomla category ID) and the article content table (containing the full German-language article text). A third table---the Joomla categories table---provides the content taxonomy.

The raw export contained all articles published from summer 2019 onward (article IDs $\geq 50{,}000$). Articles with missing or empty body text were discarded. The final dataset contains 8,939 unique articles.

\subsection{Content Taxonomy}
\label{app:taxonomy}

HardwareLuxx organizes articles in a three-level hierarchy derived from the Joomla categories table. Each category record has a \texttt{path} field that encodes the full ancestry as a slash-delimited string (e.g., \texttt{artikel/hardware/grafikkarten}). The format, editorial, and sub-category labels are read directly from the path components; no inference is required.

The two observable format categories are:
\begin{itemize}
  \item \textbf{artikel}: long-form hardware reviews, test reports, and buying guides.
  \item \textbf{news}: shorter news articles covering product announcements, industry events, and market developments.
\end{itemize}

The editorial layer contains five content categories (hardware, software, consumer-electronics, allgemein, sonstiges) under \textit{artikel} and three under \textit{news}. The sub-category layer contains 70 leaf categories in total (e.g., grafikkarten, prozessoren, mainboards under hardware). The format and editorial category labels are available as structured fields in the dataset; the sub-category label is derivable from the category path but is not exposed to the agent.

\begin{table}[h]
\centering
\small
\caption{HWL taxonomy: article counts by format and editorial category}
\label{tab:hwl_taxonomy}
\begin{tabular}{llr}
\toprule
Format & Editorial category & Articles \\
\midrule
artikel & hardware             & 1{,}371 \\
artikel & software             &   124 \\
artikel & consumer-electronics &   111 \\
artikel & sonstiges            &    18 \\
\midrule
news    & hardware             & 3{,}744 \\
news    & software             & 1{,}663 \\
news    & allgemein            & 1{,}209 \\
news    & consumer-electronics &   699 \\
\midrule
Total   &                      & 8{,}939 \\
\bottomrule
\end{tabular}
\end{table}

\subsection{WTP Calibration}
\label{app:wtp}

Willingness-to-pay is calibrated from AI-crawler traffic records. HardwareLuxx's server logs record the number of page views attributable to known AI-crawler user agents (GPTBot, ClaudeBot, PerplexityBot, and others) for each article. The article-level WTP center is:
\[
v(i) = 0.004 \times \text{AI-crawler views for article } i
\]
The coefficient $\$0.004$ per view is a calibration parameter chosen to produce a WTP distribution in a plausible per-article pay-per-crawl pricing range (cents to dollars). The query-level WTP is drawn independently for each query:
\[
v(i, q) \sim \mathcal{N}\!\left(v(i),\; \sigma^2\right), \quad \sigma = 0.001, \quad \text{truncated at } 0
\]
The small $\sigma$ captures idiosyncratic variation across requests while keeping per-query WTP close to the article-level center.

Table~\ref{tab:hwl_wtp} summarizes the resulting WTP distribution by format category. The large gap between \textit{artikel} and \textit{news} medians (\$0.060 vs.\ \$0.016) reflects that long-form hardware reviews attract substantially more AI-crawler traffic than short news items, likely because they contain the dense specifications and benchmark tables that AI systems extract.

\subsection{Train/Test Split}
\label{app:split}

Articles are split into a training set (7,210 articles, 80.7\%) and a test set (1,729 articles, 19.3\%), stratified by format category to preserve the artikel/news ratio in both sets. Stratification ensures that the format category distribution the agent encounters during training matches the distribution it is evaluated on.

Nine queries are generated per article, yielding 64,890 training queries and 15,561 test queries (80,451 total). Within each split, query arrival order is randomized across all articles so the agent does not encounter content in format-category batches.

\subsection{Query Generation}
\label{app:queries}

Each article generates nine synthetic queries. Queries are generated using a language model that reads the article title and a short excerpt, then produces plausible information-seeking requests that an AI crawler might issue. The nine queries per article vary in specificity and framing (e.g., broad topic requests, specific component questions, benchmark comparisons) to represent the distribution of crawler intents.

The agent does not observe the query text directly; it observes only the article text $t_i$ and the format category label $c(i)$. Query-level WTP is drawn independently per query as described in Section~\ref{app:wtp}.

\subsection{Information Asymmetry}
\label{app:info}

The HWL evaluation instantiates the same information asymmetry as the theoretical setting. Table~\ref{tab:info_structure} summarizes what exists in the dataset and what the agent can observe.

\begin{table}[h]
\centering
\small
\caption{Information structure (HardwareLuxx evaluation)}
\label{tab:info_structure}
\begin{tabular}{lcc}
\toprule
Data element & Exists in dataset & Agent-observable \\
\midrule
Format category $c(i)$ (artikel / news)          & \checkmark & \checkmark \\
Article title and full text $t_i$               & \checkmark & \checkmark \\
Purchase outcome $y = \mathbf{1}[p \leq v(i,q)]$ & \checkmark & \checkmark \\
Editorial / sub-category labels                  & \checkmark & $\times$ \\
AI-crawler view count                           & \checkmark & $\times$ \\
Article-level WTP center $v(i)$                 & \checkmark & $\times$ \\
Query-level WTP $v(i, q)$                       & \checkmark & $\times$ \\
\bottomrule
\end{tabular}
\end{table}

The agent has access to the same information a publisher would have in practice: the article text it already possesses and the coarse format label (review vs.\ news). All pricing-relevant structure---crawler traffic, WTP, and the finer content taxonomy---is hidden. The only feedback is binary: did the crawler purchase at the posted price or not.

\end{document}